\documentclass[twocolumn]{aastex701}
\usepackage[dvipsnames]{xcolor}

\graphicspath{{./}{Figures/}}

\begin{document}
\title{A Strongly Lensed Ultra-faint Arc at $z \approx 10$ with an F200W excess in Abell~S1063}

\author[orcid=0000-0002-3489-6381,sname='Yu']{Fujiang Yu}
\affiliation{Department of Astronomy, Tsinghua University, Beijing 100084, China}
\email{fujiang.yu@outlook.com}  

\author[0000-0001-6251-649X,sname='Li']{Mingyu Li}
\affiliation{Department of Astronomy, Tsinghua University, Beijing 100084, China}
\email{Lmytime@hotmail.com}  

\author[orcid=0000-0001-8467-6478,sname='Cai']{Zheng Cai}
\affiliation{Department of Astronomy, Tsinghua University, Beijing 100084, China}
\email[show]{zcai@mail.tsinghua.edu.cn}  

\author[orcid=0000-0002-7895-7430,sname='Guo']{Yuduo Guo}
\affiliation{College of Artificial Intelligence, Tsinghua University, Beijing 100084, China}
\affiliation{Department of Automation, Tsinghua University, Beijing 100084, China}  
\email{gyd@mail.tsinghua.edu.cn} 

\author[orcid=0009-0008-9293-8915,sname='Zhang']{Hao Zhang}
\affiliation{Department of Automation, Tsinghua University, Beijing 100084, China}
\email{haozhang_thu@mail.tsinghua.edu.cn}  

\author[orcid=0000-0001-9065-3926,sname='Diego']{Jose M. Diego}
\affiliation{Instituto de Física de Cantabria (CSIC-UC), Avda. Los Castros s/n, Santander, 39005, Spain}
\email{jdiego@ifca.es} 

\author[0009-0003-4133-0292,sname='Cai']{Sijia Cai}
\affiliation{Department of Astronomy, Tsinghua University, Beijing 100084, China}
\email{caisj23@mails.tsinghua.edu.cn}

\begin{abstract}
Strong gravitational lensing provides a powerful route to probing intrinsically faint galaxies during the first few hundred million years of cosmic history. In this Letter, we report the identification of GAR10, a highly magnified F115W-dropout galaxy at $z\approx10$ in the AbellS1063 cluster field, using deep JWST/NIRCam imaging from the GLIMPSE and GO-1840 programs. The source shows an unusually blue ultraviolet (UV) continuum and a significant F200W excess relative to adjacent bands. Under our high-magnification lensing solution, we infer a median magnification of $\mu=43^{+78}_{-20}$, corresponding to an intrinsic UV magnitude of $M_{\rm UV}\approx-15.8$. We use exploratory Prospector SED modeling to examine two physically motivated interpretations of the observed photometry. In Case I, GAR10 is described by an extremely metal-poor, continuum-dominated stellar population at $z=10.75_{-0.34}^{+0.41}$, with a blue UV slope of $\beta=-2.92\pm0.12$ and a low metallicity of $\log(Z/Z_\odot)=-3.56_{-0.85}^{+0.65}$, consistent with an extremely metal-poor or Pop III-like continuum-dominated interpretation under the adopted priors. In Case II, GAR10 is interpreted as an extremely young (1--3 Myr), high-ionization galaxy at $z=10.45_{-0.21}^{+0.11}$, in which the F200W excess is produced by intense rest-frame UV emission lines, including CIV, HeII, and CIII]. Both cases can partially reproduce the current photometry within the adopted priors, but they imply distinct ionizing sources, enrichment histories, and possible contributions to cosmic reionization. GAR10 therefore represents a rare laboratory for studying ultra-faint galaxy formation at cosmic dawn. Future JWST/NIRSpec spectroscopy will be essential to distinguishing between the steep continuum and emission-line origins of the F200W excess.
\end{abstract}


\section{Introduction}
The emergence of the first galaxies represents a pivotal epoch in cosmic history, marking the onset of star formation, chemical enrichment, and stellar mass assembly during the reionization era. While \textit{JWST} has rapidly expanded the census of galaxies at $z\gtrsim8$, probing the intrinsically faint population remains a formidable challenge \citep{2024Natur.636..332M,2025arXiv250611846N}. These faint systems are of paramount interest, as they are believed to dominate the ionizing photon budget required to reionize the Universe \citep{2024Natur.626..975A}.

Strong gravitational lensing offers a unique pathway to study this elusive population \citep{2026A&A...705A.173M}. By magnifying flux and stretching angular sizes, lensing clusters enable the spatially resolved study of sub-galactic structures in sources far below the detection threshold of blank fields \citep{2025A&A...704A..97M}. This technique is particularly powerful near critical curves, where highly magnified arcs can reveal compact clumps and detailed morphologies, providing unique insights into galaxy assembly within the first few hundred million years after the Big Bang \citep{2023ApJ...943....2W,2024Natur.632..513A}.

The physical properties of these primordial systems are intrinsically encoded in their rest-frame ultraviolet (UV) emission \citep{2016ARA&A..54..761S}. Galaxies characterized by remarkably blue UV continuum slopes (e.g., $\beta \lesssim -2.5$) typically harbor young, dust-impoverished stellar populations \citep{2024MNRAS.531..997C,2024ApJ...976..193R}. Such a blue spectral shape, particularly when associated with a lack of metal-line features or a hard ionizing spectrum, serves as a diagnostic for low-metallicity environments \citep{2025ApJ...989...46F}, providing vital constraints on the pristine conditions prevalent during the epoch of early galaxy assembly.

In this Letter, we report the discovery of a strongly gravitationally magnified arc at $z_\mathrm{phot} \approx 10$ in the Abell S1063 cluster field.
The system shows a clear F115W dropout and exhibits extreme lensing magnification of $\mu \approx 43$. In addition, its F200W flux is significantly enhanced relative to adjacent F150W, F250M, and F277W bands. With exploratory SED fitting, we examine two physically motivated interpretations: an extremely steep UV slope or intense rest-frame UV emission lines.

In the following sections, we build the case for GAR10 as a highly magnified, ultra-faint galaxy candidate at $z\approx10$. Section~\ref{sec:data} describes the JWST/NIRCam observations and data reduction, and Section~\ref{sec:source} presents the source identification, foreground-galaxy subtraction, and photometric measurements that establish GAR10 as a robust F115W-dropout source in the Abell~S1063 field. In Section~\ref{sec:analysis}, we combine photometric-redshift fitting, exploratory SED modeling, UV-continuum measurements, and a refined local lensing analysis to examine the origin of the F200W excess and to infer the intrinsic properties of the source under different magnification assumptions. In Section~\ref{sec:discussion}, we discuss two limiting physical interpretations and outline the spectroscopic tests required to distinguish between them. Finally, all results and conclusions have been summarized in section~\ref{sec:conclusion}. Throughout this work, magnitudes are given in the AB system \citep{1983ApJ...266..713O}. We adopt a flat $\Lambda$CDM cosmology with $\Omega_{\rm m}=0.3$ and $H_{0}=70~{\rm km~s^{-1}~Mpc^{-1}}$, for which $1\arcsec$ corresponds to 4.16 physical kpc at $z=10$.

\section{Observation and data reduction} \label{sec:data}

In this work, we utilize deep \textit{JWST}/NIRCam observations of the Abell~S1063 cluster field, combining data from the GLIMPSE program (GO-3293; PI: H.~Atek; \citealt{2025arXiv251107542A}) and program GO-1840 (PI: J.~\'Alvarez-M\'arquez). The combined dataset spans seven broad bands (F090W, F115W, F150W, F200W, F277W, F356W, and F444W) and four medium bands (F250M, F300M, F410M, and F480M). Primary data reduction was executed using the official \textit{JWST} Science Calibration Pipeline (v1.18.1) with the Calibration Reference Data System (CRDS) context \texttt{jwst\_1364.pmap}.

To ensure optimal image quality, we tailored the detector-level pipeline parameters within \texttt{calwebb\_detector1}, specifically refining the identification and flagging of ``snowball'' cosmic-ray artifacts prior to ramp fitting. We further mitigated detector-dependent ``wisps'' in the short-wavelength (SW) channel by applying the template-based correction described by \citet{2025jwst.rept.9225S}.

To account for the foreground contamination from the cluster, including scattered light from extended members and faint intracluster light (ICL), we implemented a custom, spatially varying background subtraction. Using \texttt{Photutils} \citep{2016ascl.soft09011B}, we estimated the 2D background by iteratively masking detected sources and employing a biweight location estimator on a sliding grid to suppress outlying flux. The final mosaics were co-added using the \texttt{calwebb\_image3} pipeline with a drizzled pixel scale of 0\farcs03~pixel$^{-1}$ and a \texttt{pixfrac} of 1.0.

We also note that the VENUS program (GO-6882, PI: S. Fujimoto) includes pointing sequence 059, which covers Abell S1063 in two medium-band filters (F210M and F300M). However, we exclude these data from our analysis due to exposure-guiding issues and obvious persistence artifacts from previous observations, which would compromise the photometric measurements.

\section{Source detection and foreground subtraction}
\label{sec:source}

\subsection{Source detection}

In this work, we first applied the ASTERIS denoising method of \citet{doi:10.1126/science.ady9404} to the JWST/NIRCam imaging of the Abell~S1063 field from program GO-3293, with the aim of improving the effective detection limit. An initial long-wavelength (LW) detection image was then constructed by coadding the ASTERIS-enhanced F277W, F356W, and F444W broad-band data.

Source detection on this detection map was performed using an iterative framework based on \texttt{SExtractor} \citep{1996A&AS..117..393B}. We progressively lowered the detection threshold from $85\sigma$ to $1.5\sigma$ using a series of manually specified, nonuniform signal-to-noise intervals, thereby expanding the source catalog from the most significant detections to fainter detections. Each source was assigned to the highest detection threshold at which it was first identified. During successive catalog merge steps, a newly detected source was retained only when its separation from all previously accepted sources exceeded $0.07\arcsec$. This procedure minimizes duplicate detections and fragmentation of bright sources while preserving distinct neighboring sources detected at lower significance.

From the resulting catalog and the corresponding multi-band photometry, we identified our primary target as a F115W-dropout source satisfying the high-redshift color-selection criteria of \citet{2025ApJ...992...63W}. By contrast, measurements from the original, non-denoised images do not place the source within the adopted color-selection region. This highlights the role of ASTERIS in improving the effective detection limit and thereby enabling the identification of high-redshift candidates even in crowded cluster environments. We hereafter refer to this candidate as GAR10 (GLIMPSE Arc at $z \approx 10$). GAR10 is additionally characterized by a prominent F200W flux excess relative to F150W, indicating an unusual rest-frame ultraviolet spectral shape.

As shown in Figure~\ref{fig:one}, the short-wavelength (SW) cutouts reveal that components $a$ and $b$ form two closely separated knots that are not fully resolved in the LW detection image. We therefore applied the same iterative source-detection framework to a detection image constructed by coadding the F150W and F200W mosaics. This SW-based detection provided independent centroids for the two knots, improving their deblending from each other and from the nearby foreground cluster member, as well as the subsequent photometric measurements of the individual GAR10 components.

\subsection{Foreground Galaxy Subtraction and Refined Photometry}

To mitigate the impact of the contamination from the bright foreground galaxy and ensure accurate photometry for the GAR10 components, we performed a foreground subtraction using a specialized two-stage iterative modeling procedure. Because accurate isophotal modeling of the foreground galaxy benefits from preserving the native noise properties of the mosaics, we performed the foreground galaxy subtraction on the original, non-denoised mosaics. The refined aperture photometry used for the photometric redshift and spectral energy distribution (SED) fitting shown in Figures~\ref{fig:eazy_redshift_estimation} and \ref{fig:Prospector_sed_results} was likewise measured from these foreground-subtracted, non-denoised mosaics.

We removed the light from the foreground galaxy using a two-stage masking and isophotal-fitting procedure. In the first stage, sources in the original image were detected at a threshold of $3\sigma$, deblended, and masked, while the segmentation region associated with the foreground galaxy was excluded from the mask and retained for modeling. The isophotal fit was initialized at the pixel position corresponding to the adopted sky coordinates of the foreground galaxy, with the center allowed to vary during the fitting. Initial estimates of the ellipticity and position angle were obtained from the second-order moments of the positive flux measured within a local region extending 25 pixels from the adopted foreground-galaxy position, after excluding masked pixels and subtracting the local median. These geometric parameters, together with an initial semi-major axis of 40 pixels, were supplied to \texttt{photutils.isophote.Ellipse} to fit a sequence of elliptical isophotes. The resulting preliminary model was subtracted from the original image, reducing the smooth foreground-galaxy light and revealing the background point sources that had previously been blended with its extended light profile.

\begin{figure*}
    \centering
    \includegraphics[width=1.0\linewidth]{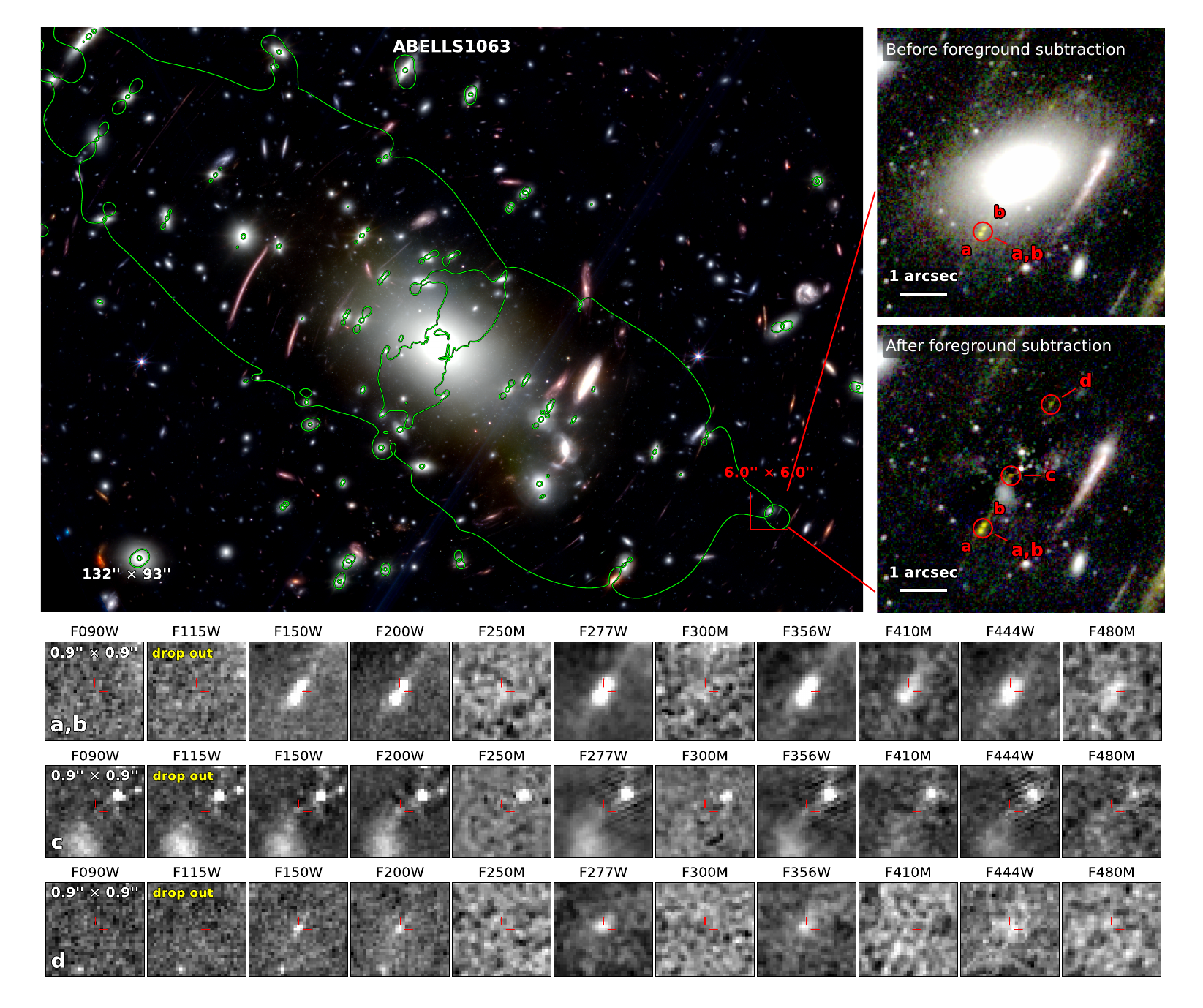}
    \caption{Multi-wavelength view of the strongly lensed ultra-faint arc at $z \approx 10$ in the Abell S1063 field. Top left: Wide-field RGB composite of the cluster core, synthesized from JWST/NIRCam filters: R (F356W + F444W), G (F200W + F277W), and B (F115W + F150W). The green lines show the critical curve at $z\approx10$ using an updated lensing model from \citet{2026arXiv260111704D}. Top right: Zoomed-in views (R: F200W, G: F150W, B: F115W) demonstrating the foreground galaxy subtraction. This procedure clearly uncovers component $d$ and the extended arc trace $c$, both of which were previously obscured. Bottom: Multi-band cutouts (from F090W to F480M) for components $ab$, $c$, and $d$, ordered by wavelength. All stamps are displayed with z-scale tied to highlight the robust F115W-dropout signature, consistent with the high-redshift interpretation of the arc structure.}
    \label{fig:one}
\end{figure*}

In the second stage, source detection was repeated in the residual image of the first-stage using a lower threshold of $2.3\sigma$. Sources detected in the residual image were added to the first-stage mask to construct a more complete composite mask. Using this updated mask, we re-fitted the isophotal model directly to the original image and subtracted the refined model to obtain the final foreground-subtracted image. This procedure substantially reduced contamination from the extended foreground light and allowed the faint lensed arc and its previously blended substructures, including components `c' and `d', to be identified. A comparison between the original and foreground-subtracted images is shown in Figure~\ref{fig:one}.

Following foreground subtraction, aperture photometry was then performed on the source cutouts, which were PSF-matched to the F480M band. The PSF models were generated using \texttt{STPSF} across all 11 bands. In this work, due to the presence of a diffuse, faint foreground galaxy near component `c' and the inherent faintness of component `d', we restrict our photometric analysis to the well-defined system `ab' to ensure accurate on-target measurements. Photometry was carried out using a circular aperture with a radius of $0\farcs2$. Photometric uncertainties were estimated by placing 20,000 random apertures within $15\arcsec$ of the target arc and calculating the standard deviation of the aperture sum.  Finally, aperture corrections were applied using factors derived from the encircled flux fraction of the F480M PSF within the aperture radius $0\farcs2$. 

\section{Analysis} 
\label{sec:analysis}
\subsection{Photometric redshift and SED fitting}
\begin{figure}
    \centering
    \includegraphics[width=1.0\linewidth]{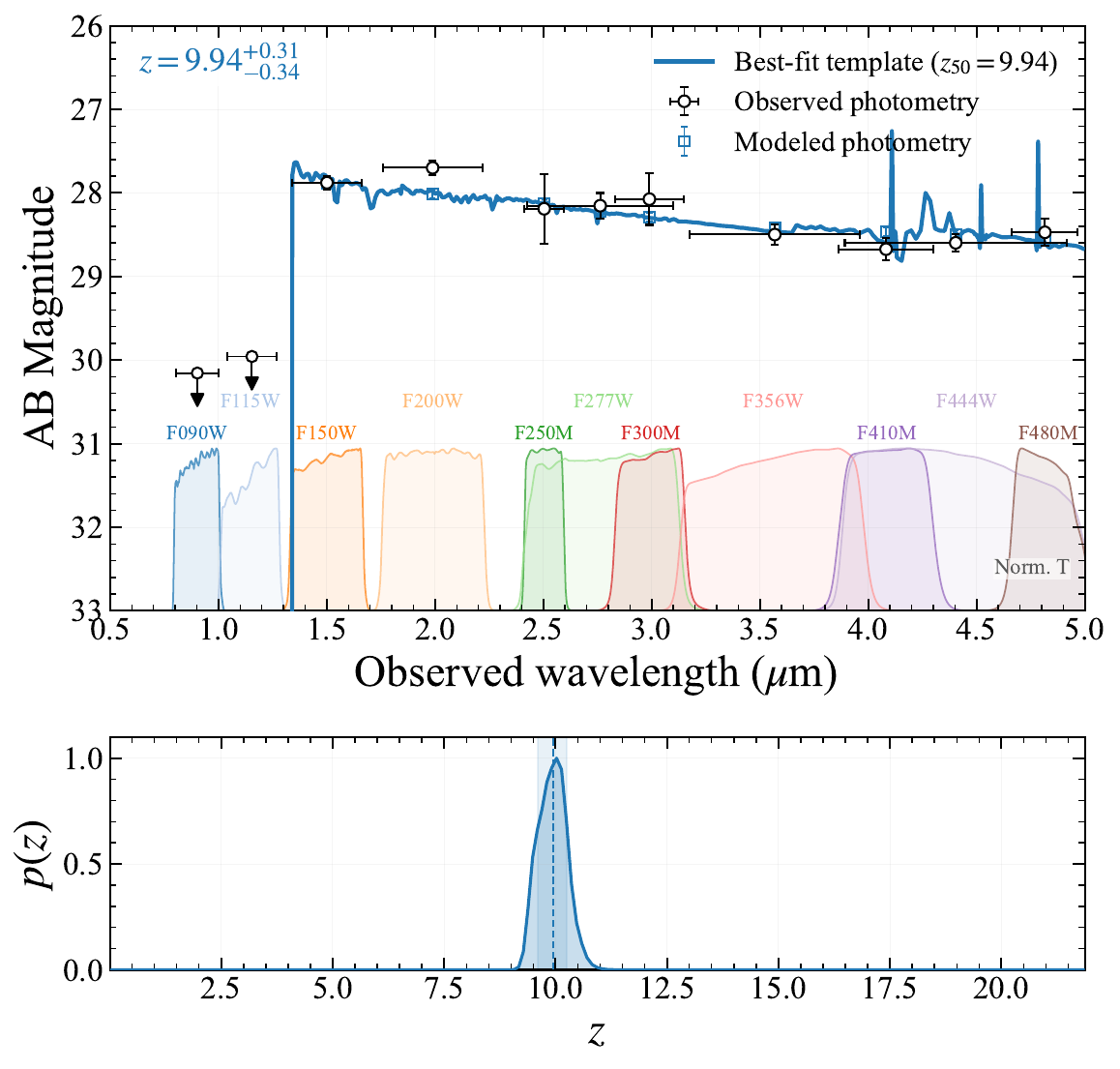}
    \caption{EAZY Photometric redshift fitting for component $ab$ of GAR10. The photometric points are measured from the foreground-subtracted, non-denoised JWST/NIRCam mosaics. The best-fit template, shown by the blue line, yields a high-redshift solution with a sharp Lyman alpha break and demonstrates that F200W exhibits a notable flux excess relative to F150W.}
    \label{fig:eazy_redshift_estimation}
\end{figure}

To derive primary constraints on the photometric redshift, we performed spectral energy distribution (SED) fitting using the \texttt{EAZY} code \citep{2008ApJ...686.1503B} on the photometry measured from the foreground-subtracted images. We adopted the standard galaxy template set from \citet{2010ApJ...712..833C, 2006AJ....132..926C, 2010ApJ...719.1168E, 2008ApJ...686.1503B}. As shown in Figure~\ref{fig:eazy_redshift_estimation}, the best-fit template presents a sharp Lyman alpha break, characterized by a complete dropout in the F115W band ($<2\sigma$) and a robust detection in F150W. This strong spectral break provides a well-constrained redshift solution of $z=9.94^{+0.31}_{-0.34}$, effectively ruling out low-redshift solutions that would otherwise predict significant flux in the bluer NIRCam filters.

While the template-fitting approach with \texttt{EAZY} provides an initial redshift estimate, the adopted template library does not adequately reproduce the significant F200W flux excess. The physical origin of this discrepancy cannot be determined from the \texttt{EAZY} fit alone. We therefore employ physically motivated SED modeling to investigate whether the F200W excess can be explained by an extremely blue stellar continuum or by strong rest-frame UV emission lines.

The observed F200W excess in GAR10 is intrinsically coupled to the redshift estimate, because the redshift determines how the Lyman-$\alpha$ break and the underlying stellar continuum contribute to the flux measured in the F150W bandpass. To explore this degeneracy, we performed SED modeling with the Bayesian inference framework \texttt{Prospector} \citep{2021ApJS..254...22J}, adopting two sets of physically motivated priors that are intended to probe two limiting scenarios.

In Case~I (Ultra-blue continuum dominant model), we examine a continuum-driven interpretation by adopting a broad redshift prior ($9.8 < z < 14.0$) and allowing the stellar metallicity to extend down to $\log(Z/Z_\odot) = -5.0$. This setup permits extremely blue ultraviolet (UV) continuum slopes ($\beta$), as expected for massive, metal-poor stellar populations, while allowing the position of the Lyman alpha break to vary in order to reproduce the observed photometry.

In Case~II (EELG model), we explore a line-driven interpretation by restricting the redshift to a narrower interval ($9.8 < z < 10.6$), such that high-ionization rest-frame UV emission lines, including \ion{He}{2}~$\lambda1640$ and \ion{C}{3}]\,$\lambda1908$, fall within the F200W filter. This model assumes a very young stellar population ($t_{\text{age}} \in [1, 3]$ Myr) and a high ionization parameter ($\log U \in [-3.0, -1.0]$), thereby enhancing the nebular contribution to the broadband flux.

For both cases, we assume a dust-free environment ($A_V = 0$) and a burst-like star formation history ($\tau < 0.01$ Gyr). We emphasize that these models should be regarded as exploratory rather than definitive. They are designed to test whether the observed F200W excess can be plausibly explained by either an extremely blue, metal-poor stellar continuum or by intense rest-frame UV nebular emission, rather than to uniquely determine the physical nature of the source.

\subsection{SED Results and UV Properties}

Our \texttt{Prospector} analysis identifies two distinct classes of solutions that reproduce the observed photometry, particularly the prominent F200W flux excess. The physical properties derived from these solutions are summarized below and in Table~\ref{tab:results}.

We derived the UV continuum slope $\beta$ by performing a power-law fit ($f_\lambda \propto \lambda^\beta$) to the best-fit \texttt{Prospector} spectrum. To ensure a reliable measurement of the stellar continuum, we adopted a ``line-clean'' fitting procedure over the rest-frame wavelength range of $1450$--$2600$~\AA, masking regions within $\pm40$~\AA\ of known strong emission lines, including \ion{C}{4}~$\lambda1549$, \ion{He}{2}~$\lambda1640$, and \ion{C}{3}]\,$\lambda1908$. An iterative sigma-clipping algorithm ($2.5\sigma$) was applied during the fit to further remove remaining non-continuum features, ensuring that the derived $\beta$ reflects the underlying continuum.

The absolute UV magnitude, $M_{\rm UV}$, was estimated by integrating the best-fit model spectra through a synthetic 100~\AA-wide top-hat filter centered at a rest-frame wavelength of 1500~\AA\ and covering 1450--1550~\AA. The median values and associated $1\sigma$ uncertainties were obtained from the posterior distribution by re-evaluating the rest-frame SEDs across the MCMC/nested-sampling chains.

\begin{figure*}[t]
\centering
\includegraphics[width=1.0\textwidth]{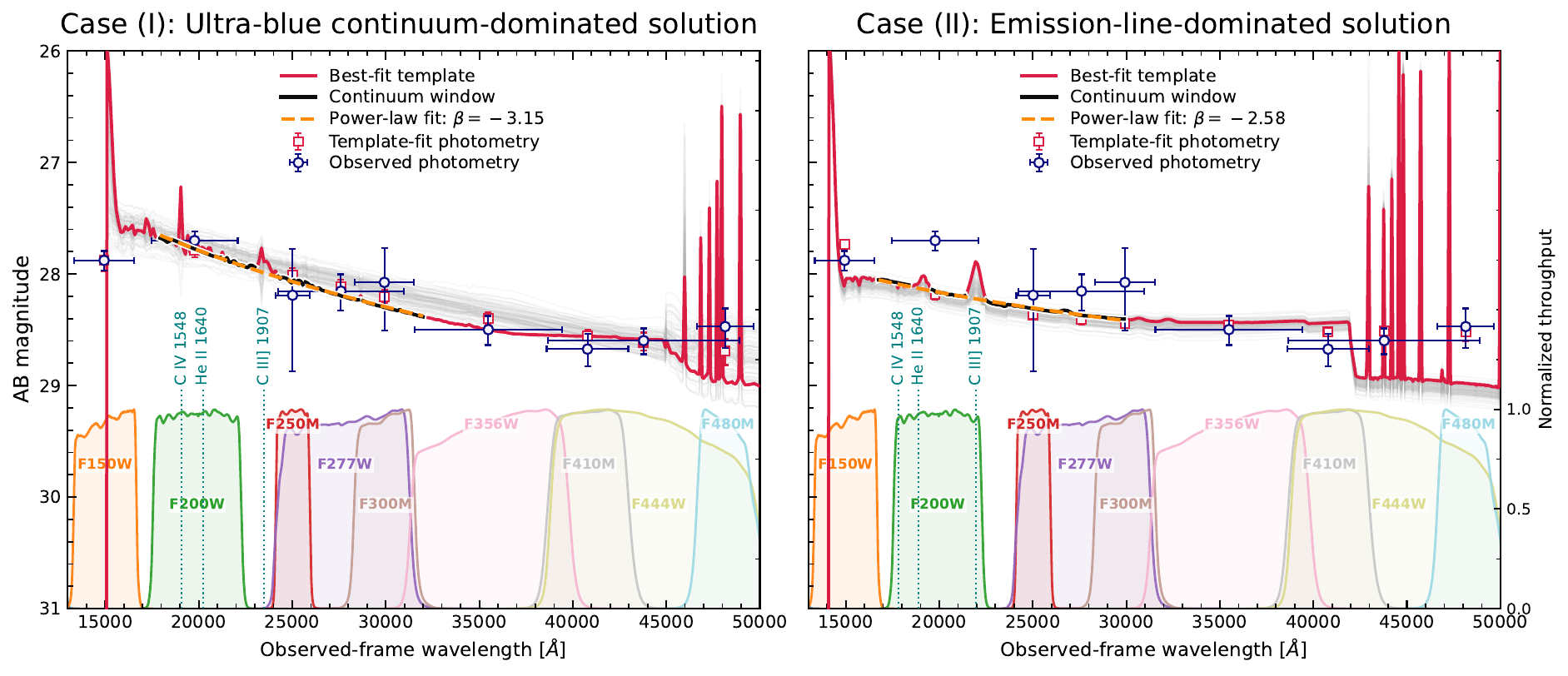}
\caption{
Best-fit \texttt{Prospector} SED modeling for component $ab$ of GAR10 under two exploratory limiting scenarios.
\textbf{Left panel (Case~I):}
A continuum-dominated, extremely metal-poor solution.
The red curve shows a representative best-fit realization at $z_{\rm best}\approx11.33$.
All physical quantities quoted for this panel are measured from this best-fit model spectrum rather than from the marginalized posterior medians.
A line-clean power-law fit to the best-fit spectrum gives an exceptionally blue UV continuum slope of $\beta_{\rm best}=-3.15$, while the best-fit stellar metallicity reaches $\left[\log_{10}\left(Z/Z_\odot\right)\right]_{\rm best} \approx -4.9$.
These best-fit values differ from the marginalized posterior medians reported in Table~\ref{tab:results} ($z = 10.75^{+0.41}_{-0.34}$, $\beta=-2.92\pm0.12$, and $\log(Z/Z_\odot)=-3.56_{-0.85}^{+0.65}$), because the Case~I posterior is non-Gaussian and extends toward the edge of the adopted low-metallicity, dust-free parameter space.
We therefore interpret this solution as an illustrative limiting case in which the F200W flux is dominated primarily by an ultra-blue stellar continuum, with negligible nebular-line contribution.
\textbf{Right panel (Case~II):}
A lower-redshift, line-driven solution.
The red curve shows the representative best-fit realization at $z_{\rm best}\approx10.52$.
As in the left panel, all quoted quantities are measured from the best-fit model spectrum.
A line-clean power-law fit to this best-fit spectrum gives a more moderate UV continuum slope of $\beta_{\rm best}=-2.58$.
This solution corresponds to an extremely young burst age of $1$--$3$~Myr in the adopted Case~II model setup.
The apparent inverted Balmer break feature is therefore naturally associated with a very young, high-ionization stellar population with strong nebular emission.
In this scenario, the F200W excess is reproduced by strong rest-frame UV nebular emission, primarily \ion{He}{2}~$\lambda1640$ and the \ion{C}{3}]\,$\lambda1908$, with a possible contribution from \ion{C}{4}~$\lambda1549$.
The F200W flux exceeds the modeled line-free continuum by approximately $4.4\sigma$. Based on this flux excess and the rectangular width of the F200W filter, we infer a rest-frame equivalent width of $EW_{\rm rest,F200W}\approx211$~\AA.
In both panels, the red curve denotes the best-fit model spectrum, the gray shaded regions show the posterior model distribution, and the navy points indicate the observed JWST/NIRCam photometry.
The NIRCam filter-throughput curves are shown in the lower sub-panels.
}
\label{fig:Prospector_sed_results}
\end{figure*}

Case~I corresponds to a continuum-dominated, extremely metal-poor solution whose marginalized posterior medians are listed in Table~\ref{tab:results}:$z=10.75_{-0.34}^{+0.41}$, $\beta=-2.92\pm0.12$, and $\log(Z/Z_\odot)=-3.56_{-0.85}^{+0.65}$. However, the representative best-fit/MAP realization shown in Figure~\ref{fig:Prospector_sed_results} lies at the high-redshift, low-metallicity edge of the allowed parameter space, with $z_{\rm best}\approx11.33$, $\beta_{\rm best}\approx-3.15$, and $\log(Z/Z_\odot)_{\rm best}\approx-4.9$. This offset between the marginalized values and the best-fit realization arises because the Case~I posterior is non-Gaussian and extends toward the adopted prior boundaries, particularly the very low-metallicity, dust-free limit. We therefore interpret the best-fit spectrum as an illustrative limiting case in which the F200W flux is dominated by an ultra-blue stellar continuum, with negligible contribution from nebular lines.

Case~II favors a slightly lower-redshift solution, with a marginalized redshift of $z=10.45_{-0.21}^{+0.11}$ and a representative best-fit value of $z_{\rm best}\approx10.52$. In this scenario, the SED exhibits a more moderate UV slope of $\beta=-2.56\pm0.02$. The F200W flux excess is primarily modeled as a suite of high-ionization nebular emission lines, including \ion{He}{2}~$\lambda1640$ and the \ion{C}{3}]\,$\lambda1908$, originating from a system at an early stage of chemical enrichment, with $\log(Z/Z_\odot)\approx-1.55$. Under the adopted model setup, this solution corresponds to an extremely young burst age of $1$--$3$~Myr and a high-ionization stellar population, consistent with the apparent inverted Balmer-break-like spectral shape in the best-fit model.

As shown in Figure~\ref{fig:Prospector_sed_results}, the best-fit Case~II model produces rest-frame UV emission lines within the F200W band, but these modeled lines are not sufficiently strong to fully account for the observed F200W flux excess. We therefore estimate the equivalent width required to reproduce the observed excess relative to the inferred underlying continuum. To quantify this excess, we calculated a broadband-inferred equivalent width. The continuum level within the F200W filter was determined from the line-free power-law continuum model described above. The observed-frame equivalent width was calculated as
\begin{equation}
EW_{\rm obs} = W_{\rm rect} \times
\frac{f_{\rm obs}-f_{\rm cont}}{f_{\rm cont}},
\end{equation}
where $W_{\rm rect}=4609$~\AA\ is the rectangular width of the F200W filter, $f_{\rm obs}$ is the observed photometric flux density, and $f_{\rm cont}$ is the predicted continuum flux density from the power-law fit. The rest-frame equivalent width was then derived as
$EW_{\rm rest}=EW_{\rm obs}/(1+z)$.
This calculation yields a broadband-inferred rest-frame equivalent width of $EW_{\rm rest,F200W}\approx211$~\AA\ for Case~II.

We note that, although both models provide statistically reliable fits to the current multi-band photometry, the degeneracy between a continuum-driven and a line-driven SED underscores the extreme nature of GAR10. Given the limitations of broadband photometry in distinguishing between these physical interpretations, future \textit{JWST}/NIRSpec spectroscopy will be essential to break this degeneracy and determine the physical origin of the F200W flux excess.

\subsection{Lensing Magnification and Intrinsic Luminosity}

As illustrated in Figure~\ref{fig:one}, the primary target components $a$ and $b$, together with the elongated structure labeled trace $c$, lie close to the core of a foreground cluster member. The clumpy morphology and tangentially stretched configuration are consistent with a highly magnified lensed arc. However, accurately determining the magnification factor ($\mu$) and the expected counterimage configuration remains challenging, because the source lies in a region where the lensing solution is highly sensitive to the local mass distribution of the foreground member galaxy and its coupling to the cluster-scale potential.

\begin{figure}
    \centering
    \includegraphics[width=1.0\linewidth]{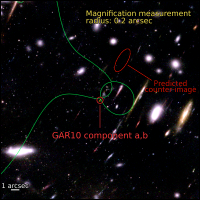}
    \caption{Critical curves from the gravitational-lensing model under Scenario 1 are shown, passing directly through the target knots ``ab''. The red ellipse at upper right indicates the predicted positional range of the counterimage, which lies in a low-magnification region. The magnification factor is estimated from the magnification map within a $0\farcs2$-radius aperture centered on component ``ab''. We obtain a median magnification of $\mu = 43$, with the 16th and 84th percentiles giving a $1\sigma$ range of $[23, 121]$ and a mean value of $\mu \approx 99$.
}
    \label{fig:Critical_Curve}
\end{figure}

We therefore consider two possible lensing interpretations, distinguished primarily by the location of the critical curve relative to components $a$ and $b$. In Scenario~1, the critical curve passes through the region between the two knots, placing component $ab$ in an extreme-magnification regime produced by the combined effects of the cluster-scale potential and the local galaxy-scale perturbation associated with the nearby cluster member. In this configuration, the predicted counterimage is expected to lie within the region indicated in Figure~\ref{fig:Critical_Curve}. Because this region has a relatively low magnification, the counterimage is expected to be too faint to be reliably identified in the current imaging data. In Scenario~2, the nearby cluster member is assigned a lower mass, such that the critical curve does not pass through the $ab$ structure and the system instead experiences a more moderate magnification of $\mu\sim10$ \citep{2026arXiv260111704D}. The nature of the elongated trace $c$ remains uncertain in both scenarios and is therefore not used to distinguish between the two lensing interpretations.

For Scenario 1, we performed additional local refinements to the lensing model originally presented by \citet{2026arXiv260111704D}. Starting from their cluster-scale mass model, we explored a modified configuration in which the mass of the foreground cluster member galaxy is increased, shifting the cluster critical curve closer to the GAR10 double-knot structure until it produces a pair of images near the location of ab. This configuration captures the combined lensing effect of the cluster-scale potential and the local deflector, and can produce local magnifications of order $\mu \sim 100$ near the critical curve. Under this refined configuration, the mass assigned to the foreground member galaxy is approximately $3 \times 10^{11}\,M_{\odot}$. This value is likely elevated because the local galaxy component partly compensates for uncertainties in the macro-model magnification in this region; the physical stellar mass of the member galaxy is therefore expected to be lower, plausibly below $2 \times 10^{11}\,M_{\odot}$.

Because the magnification field varies rapidly across the target region, we do not adopt the peak magnification directly. Instead, we estimate the effective magnification for the photometric aperture by measuring the magnification distribution from the refined model map within a circular aperture of radius $0\farcs2$ centered on components $a$ and $b$. This yields a median magnification of $\mu_{\rm med}=43$, with the 16th--84th percentile range spanning $\mu=23$--$121$ and a mean value of $\langle\mu\rangle \approx 99$. The difference between the median and mean reflects the highly skewed magnification distribution expected near a critical curve. We adopt the median value as our fiducial correction for the intrinsic luminosity and stellar mass, while using the percentile range to represent the lensing uncertainty. The corresponding magnification corrected physical properties are summarized in Table~\ref{tab:results}.

\begin{deluxetable*}{lcc}
\tablecaption{Physical Properties of the Target Source \label{tab:results}}
\tablecolumns{3}
\tablewidth{0pt}
\tablehead{
\colhead{Parameter} & \colhead{Case~I} & \colhead{Case~II}
}
\startdata
R.A. (deg) & \multicolumn{2}{c}{342.1630271} \\
Decl. (deg) & \multicolumn{2}{c}{-44.5385197} \\
Redshift ($z$) & $10.75_{-0.34}^{+0.41}$ & $10.45_{-0.21}^{+0.11}$ \\
Best-fit redshift ($z_{\rm best}$) & $11.33$ & $10.52$ \\
\hline
$M_{UV, \text{obs}}$ (mag) & $-19.89 \pm 0.09$ & $-19.42 \pm 0.06$ \\
$M_{UV, \text{corr}}$ (mag)\tablenotemark{a} & $-15.80_{-0.70}^{+1.11}$ & $-15.33_{-0.70}^{+1.11}$ \\
UV slope ($\beta$) & $-2.92 \pm 0.12$ & $-2.56 \pm 0.02$ \\
$\log(M_\star/M_\odot)_{\text{obs}}$ & $8.46_{-0.30}^{+0.25}$ & $7.42_{-0.04}^{+0.03}$ \\
$\log(M_\star/M_\odot)_{\text{corr}}$\tablenotemark{a} & $6.82_{-0.45}^{+0.28}$ & $5.78_{-0.45}^{+0.28}$ \\
Rest-frame EW (\AA) & \nodata & 211 \\
$\log(Z/Z_\odot)$ & $-3.56_{-0.85}^{+0.65}$ & $-1.55_{-1.03}^{+0.44}$ \\
\enddata
\tablecomments{Case~I and Case~II denote the pristine/Pop III-like continuum scenario and the extreme emission-line galaxy (EELG) scenario, respectively. The listed equivalent width is a broadband-inferred rest-frame F200W excess and is applicable only to the line-driven Case~II interpretation.}
\tablenotetext{a}{Corrected for a median magnification of $\mu = 43$. The errors of the corrected $M_{UV}$ and stellar mass were estimated based on the $1\sigma$ magnification range from $\mu = 23$ to $121$.}
\end{deluxetable*}

The magnification corrected quantities listed in Table~\ref{tab:results} are derived under our fiducial high-magnification Scenario~1, using $\mu_{\rm med}=43$. Under this assumption, the physical parameters inferred from the \texttt{Prospector} SED modeling suggest two distinct limiting interpretations for component $ab$. In Case~I ($z \approx 10.75$), the SED is reproduced by an extremely metal-poor stellar population, with $\log(Z/Z_\odot)= -3.56_{-0.85}^{+0.65}$ and a very blue UV continuum slope of $\beta=-2.92\pm0.12$. Together with the inferred faint intrinsic magnitude, $M_{\rm UV}\approx -15.8$, this solution is consistent with a Population~III-like or extremely metal-poor continuum-dominated interpretation. We emphasize, however, that this should be regarded as a physically motivated limiting case rather than a unique identification of a pristine stellar population. In Case~II ($z \approx 10.45$), the F200W excess is instead attributed to strong rest-frame UV nebular emission, including \ion{He}{2}~$\lambda1640$, \ion{C}{4}~$\lambda1549$, and \ion{C}{3}]\,$\lambda1908$. The low inferred stellar mass, $\log(M_\star/M_\odot)\approx5.78$, and the required high-ionization emission suggest a very young, compact, bursty system with an unusually hard radiation field.

Alternatively, in Scenario~2, the observed morphology may correspond to a more moderately magnified system in which the trace $c$ is an intrinsic structure, such as a tidal feature or close companion, rather than a separate lensed counterimage \citep{2026arXiv260106015Y}. In this case, the magnification would be closer to $\mu\sim10$, as in the original lensing calibration of \citet{2026arXiv260111704D}, and the inferred intrinsic properties would shift accordingly. Relative to the fiducial $\mu_{\rm med}=43$ correction, adopting $\mu\sim10$ would make the source intrinsically brighter by $\approx1.6$~mag and more massive by $\approx0.64$~dex. Nevertheless, the qualitative SED degeneracy between a continuum-driven and a line-driven origin of the F200W excess would remain. Given the current lack of spectroscopic constraints, we therefore retain both lensing interpretations and treat the intrinsic luminosity and stellar mass as magnification-dependent quantities.

\section{Discussion}
\label{sec:discussion}

The discovery of \object{GAR10} provides a rare opportunity to study an intrinsically faint, strongly lensed galaxy candidate at $z \approx 10$ near the reionization frontier. Its F115W-dropout signature, unusually blue UV continuum, and prominent F200W flux excess point to an extreme radiation field, but the current photometry does not uniquely determine the underlying physical origin. Instead, our analysis highlights a degeneracy between two physically motivated limiting scenarios: an extremely metal-poor, continuum-dominated stellar population and a very young, high-ionization emission-line galaxy.

\subsection{Nature of the Ionizing Source: Pop III-like Continuum or High-Ionization Nebular Emission}

The two SED solutions imply distinct interpretations for the ionizing source in \object{GAR10}. In Case~I, the extremely blue UV slope ($\beta \approx -2.92$) and low inferred metallicity ($\log(Z/Z_\odot) \approx -3.56$) are consistent with expectations for a very young, extremely metal-poor, possibly Population~III-like stellar population \citep[e.g.,][]{2003A&A...397..527S}. Similar blue UV continuum has been found in JWST/NIRCam-selected galaxies at $z\simeq7$--$11$ and interpreted as possible signatures of low-metallicity massive stars and hard ionizing spectra \citep{2022ApJ...941..153T}. Under this interpretation, the F200W excess is driven primarily by an ultra-blue stellar continuum. The intrinsic faintness inferred under the high-magnification lensing solution ($M_{\rm UV} \approx -15.8$) further suggests that \object{GAR10} may probe a low-mass system in an early stage of chemical enrichment, close to the regime where strong lensing is required to access chemically primitive stellar populations \citep{2023A&A...678A.173V,2025ApJ...989...46F}. We emphasize, however, that this interpretation should be regarded as a limiting case rather than a unique identification of a pristine Population~III system, because the inferred properties depend on the adopted priors and the ability of the stellar population models to represent such extreme conditions.

In Case~II, the F200W excess is instead interpreted as being produced by intense rest-frame UV emission. In this scenario, high-ionization lines such as \ion{He}{2}~$\lambda1640$, \ion{C}{4}~$\lambda1549$, and \ion{C}{3}]\,$\lambda1908$ contribute significantly to the F200W flux, while the UV continuum slope is moderately flatter ($\beta \approx -2.56$). Similar high-ionization UV-line spectra have been observed with \textit{JWST}/NIRSpec in galaxies at comparable or earlier cosmic epochs, including GHZ2/GLASS-z12 at $z=12.34$ \citep{2024ApJ...972..143C}. The inferred low stellar mass and young age suggest a compact, bursty system with a hard ionizing spectrum and a high ionization parameter. Such conditions are commonly associated with low-metallicity, high-ionization galaxies, although the presence of metal lines such as \ion{C}{4} and \ion{C}{3}] would imply a nonzero level of chemical enrichment \citep{2019MNRAS.488.3492S,2024MNRAS.529.3301T}. In our line-driven fit, ordinary UV metal-line emission alone does not fully account for the observed F200W excess; if the excess is at least partly line-dominated, an additional contribution from strong \ion{He}{2}~$\lambda1640$ may be applicable. Strong nebular \ion{He}{2} emission is a key diagnostic of an unusually hard ionizing spectrum and has been discussed as a possible signature of very hot, metal-poor or Population~III-like massive stars \citep{2023MNRAS.525.5328T,2024A&A...687A..67M}. Such a source would therefore represent an extreme high-redshift analog of UV-line-dominated emission-line galaxies, but in a substantially lower-mass regime than typically accessible without strong lensing.

\subsection{Implications and Future Spectroscopic Tests}

Although the physical origin of the F200W excess remains ambiguous, both interpretations suggest that \object{GAR10} hosts an unusually hard radiation field in an intrinsically faint candidate of a low-mass galaxy. Under the high-magnification solution adopted in Table~\ref{tab:results}, the inferred stellar mass is only $M_\star \sim 10^{5.8}$--$10^{6.8}\,M_\odot$, placing \object{GAR10} close to the low-mass frontier currently accessible at $z \gtrsim 10$. While a single lensed object cannot constrain the global ionizing photon budget, \object{GAR10} provides a glimpse of the faint dwarf-galaxy population that is expected to contribute substantially to cosmic reionization \citep[e.g.,][]{2024Natur.626..975A}. Its combination of low stellar mass, young inferred age, and hard radiation field suggests that very low-mass galaxies may already host efficient ionizing sources within the first few hundred million years of cosmic history.

There are two important caveats to this interpretation. First, the intrinsic luminosity and stellar mass estimates depend directly on the adopted lensing magnification. Our fiducial value of $\mu = 43^{+78}_{-20}$ is derived from a refined local lens model in which the critical curve is shifted close to the foreground member galaxy. Because this configuration depends sensitively on the mass assigned to the member galaxy, which can reach $\sim 3\times10^{11}\,M_\odot$ in the refined model, the magnification remains model-dependent. Second, the SED-based physical parameters are sensitive to the adopted priors and model assumptions. This is particularly relevant for Case~I, where the best-fit solution approaches the low-metallicity and minimal-attenuation boundary of the allowed parameter space. The two SED solutions should therefore be interpreted as physically motivated limiting scenarios rather than unique determinations of the stellar population and ionizing source.

The present interpretation is ultimately limited by the degeneracies inherent in broadband photometry. Deep \textit{JWST}/NIRSpec spectroscopy will be required to test the origin of the F200W excess directly. A detection of \ion{He}{2}~$\lambda1640$ would constrain the hardness of the ionizing spectrum, while measurements or upper limits in \ion{C}{4}~$\lambda1549$, \ion{O}{3}], and \ion{C}{3}] would constrain the level of chemical enrichment and the contribution of high-ionization nebula emission. Line ratios, equivalent widths, and line profiles would further distinguish between an extremely blue stellar continuum and a line-dominated interpretation. Such observations will determine whether \object{GAR10} is best understood as an extremely metal-poor, continuum-dominated source or as a very low-mass, high-ionization galaxy at $z\approx10$.

\section{Summary}
\label{sec:conclusion}

In this work, we have presented the photometric and lensing analysis of GAR10, a strongly magnified high redshift galaxy candidate at $z\approx10$ in the Abell~S1063 cluster field. The combination of its intrinsically faint UV luminosity, low inferred stellar mass, extremely blue continuum, and prominent F200W excess indicates that GAR10 is an unusually young system with a hard radiation field. Although the current photometry and SED fitting do not yet provide sufficient constraints to establish a definitive physical origin for the F200W excess, our analysis identifies two physically motivated limiting interpretations.

The main results are summarized as follows

\begin{enumerate}
\item Under our fiducial high-magnification lensing model, the primary source components have a median magnification of $\mu = 43_{-20}^{+78}$. The corresponding intrinsic properties, $M_{\rm UV}\approx -15.8$ and $M_\star \sim 10^{5.8}$--$10^{6.8}\,M_\odot$, place GAR10 in the regime of extremely low-mass dwarf galaxies that would otherwise remain below the detection limits of current observations at $z\approx10$.

\item Exploratory SED modeling with \texttt{Prospector} reveals a significant degeneracy in the physical origin of the F200W excess, motivating two representative interpretations.

\item In Case~I, the photometry is reproduced by a continuum-dominated, extremely metal-poor stellar population at $z = 10.75_{-0.34}^{+0.41}$, with $\log(Z/Z_\odot)\approx -3.56$ and a very blue UV continuum slope of $\beta\approx -2.92$. This solution is consistent with a very young system at an early stage of chemical enrichment, potentially approaching Population~III-like conditions.

\item In Case~II, the F200W excess is instead explained by an extreme emission-line galaxy at $z = 10.45_{-0.21}^{+0.11}$. Strong rest-frame UV nebular lines, including \ion{He}{2}~$\lambda1640$, \ion{C}{4}~$\lambda1549$, and the \ion{C}{3}]\,$\lambda1908$, contribute substantially to the broadband flux, indicating a very young ($1$--$3$~Myr), compact, high-ionization star-forming system.

\end{enumerate}

In either interpretation, GAR10 offers a glimpse of the low-mass, metal-poor dwarf-galaxy population at $z\approx 10$ that remains largely beyond the reach of current blank-field JWST observations. Such intrinsically faint systems are expected to contribute substantially to the ionizing photon budget during cosmic reionization, yet their stellar populations, chemical enrichment, and ionizing radiation fields remain poorly constrained observationally. By bringing one such system into view, strong gravitational lensing provides a rare opportunity to probe the physical conditions of galaxies in this otherwise inaccessible regime. Deep \textit{JWST}/NIRSpec spectroscopy will be essential for determining whether the F200W excess is driven primarily by an extremely blue, metal-poor stellar continuum or by high-ionization UV emission lines.

\begin{acknowledgments}
We thank Fengwu Sun, Xiaojing Lin, Yunjing Wu, Xiaoyang Wei, and Shuaiyi Li for helpful discussions on technical aspects of this work and its future prospects. This work is supported by National Key R\&D Program of China (grant no. 2023YFA1605600), This research is supported by National Natural Science Foundation of China (\#12525303) and Tsinghua University Initiative Scientific Research Program.  This work is funded by New Cornerstone Science Foundation through the XPLORER PRIZE.

This work is based on observations made with the NASA/ESA/CSA 
James Webb Space Telescope. The data were obtained from the 
Mikulski Archive for Space Telescopes (MAST) at the Space Telescope 
Science Institute, which is operated by the Association of Universities for Research in Astronomy, Inc., under NASA contract NAS 5-03127 for JWST. This work makes use of JWST/NIRCam imaging data from programs GO-3293 and GO-1840. All the JWST raw data used in this Letter can be found in MAST: doi:\href{https://doi.org/10.17909/z58f-0330}{10.17909/z58f-0330}. We also acknowledge the VENUS program GO-6882, whose observations of the Abell~S1063 field were inspected in this work. We thank the PIs and observing teams of these programs for designing and obtaining the observations and for making the data available with no exclusive-access period.

\end{acknowledgments}

\facilities{JWST(NIRCam)}
\software{
astropy \citep{2022ApJ...935..167A},
photutils \citep{2016ascl.soft09011B},
Source Extractor \citep{1996A&AS..117..393B},
EAZY \citep{2008ApJ...686.1503B},
Prospector \citep{2021ApJS..254...22J},
}

\bibliography{sample701}{}

@ARTICLE{1996A&AS..117..393B,
       author = {{Bertin}, E. and {Arnouts}, S.},
        title = "{SExtractor: Software for source extraction.}",
      journal = {\aaps},
         year = 1996,
        month = jun,
       volume = {117},
        pages = {393-404},
          doi = {10.1051/aas:1996164},
       adsurl = {https://ui.adsabs.harvard.edu/abs/1996A&AS..117..393B}
}

@ARTICLE{2025arXiv250611846N,
       author = {{Nakajima}, Kimihiko and {Ouchi}, Masami and {Harikane}, Yuichi and {Vanzella}, Eros and {Ono}, Yoshiaki and {Isobe}, Yuki and {Nishigaki}, Moka and {Tsujimoto}, Takuji and {Nakamura}, Fumitaka and {Xu}, Yi and {Umeda}, Hiroya and {Zhang}, Yechi},
        title = "{An Ultra-Faint, Chemically Primitive Galaxy Forming in the Reionization Era}",
      journal = {arXiv e-prints},
         year = 2025,
        month = jun,
          eid = {arXiv:2506.11846},
        pages = {arXiv:2506.11846},
          doi = {10.48550/arXiv.2506.11846},
archivePrefix = {arXiv},
       eprint = {2506.11846},
 primaryClass = {astro-ph.GA},
       adsurl = {https://ui.adsabs.harvard.edu/abs/2025arXiv250611846N}
}

@ARTICLE{2024Natur.636..332M,
       author = {{Mowla}, Lamiya and {Iyer}, Kartheik and {Asada}, Yoshihisa and {Desprez}, Guillaume and {Tan}, Vivian Yun Yan and {Martis}, Nicholas and {Sarrouh}, Ghassan and {Strait}, Victoria and {Abraham}, Roberto and {Brada{\v{c}}}, Maru{\v{s}}a and {Brammer}, Gabriel and {Muzzin}, Adam and {Pacifici}, Camilla and {Ravindranath}, Swara and {Sawicki}, Marcin and {Willott}, Chris and {Estrada-Carpenter}, Vince and {Jahan}, Nusrath and {Noirot}, Ga{\"e}l and {Matharu}, Jasleen and {Rihtar{\v{s}}i{\v{c}}}, Gregor and {Zabl}, Johannes},
        title = "{Formation of a low-mass galaxy from star clusters in a 600-million-year-old Universe}",
      journal = {\nat},
         year = 2024,
        month = dec,
       volume = {636},
       number = {8042},
        pages = {332-336},
          doi = {10.1038/s41586-024-08293-0},
archivePrefix = {arXiv},
       eprint = {2402.08696},
 primaryClass = {astro-ph.GA},
       adsurl = {https://ui.adsabs.harvard.edu/abs/2024Natur.636..332M}
}

@ARTICLE{2024Natur.626..975A,
       author = {{Atek}, Hakim and {Labb{\'e}}, Ivo and {Furtak}, Lukas J. and {Chemerynska}, Iryna and {Fujimoto}, Seiji and {Setton}, David J. and {Miller}, Tim B. and {Oesch}, Pascal and {Bezanson}, Rachel and {Price}, Sedona H. and {Dayal}, Pratika and {Zitrin}, Adi and {Kokorev}, Vasily and {Weaver}, John R. and {Brammer}, Gabriel and {Dokkum}, Pieter van and {Williams}, Christina C. and {Cutler}, Sam E. and {Feldmann}, Robert and {Fudamoto}, Yoshinobu and {Greene}, Jenny E. and {Leja}, Joel and {Maseda}, Michael V. and {Muzzin}, Adam and {Pan}, Richard and {Papovich}, Casey and {Nelson}, Erica J. and {Nanayakkara}, Themiya and {Stark}, Daniel P. and {Stefanon}, Mauro and {Suess}, Katherine A. and {Wang}, Bingjie and {Whitaker}, Katherine E.},
        title = "{Most of the photons that reionized the Universe came from dwarf galaxies}",
      journal = {\nat},
         year = 2024,
        month = feb,
       volume = {626},
       number = {8001},
        pages = {975-978},
          doi = {10.1038/s41586-024-07043-6},
archivePrefix = {arXiv},
       eprint = {2308.08540},
 primaryClass = {astro-ph.GA},
       adsurl = {https://ui.adsabs.harvard.edu/abs/2024Natur.626..975A}
}

@ARTICLE{2026A&A...705A.173M,
       author = {{Messa}, M. and {Vanzella}, E. and {Loiacono}, F. and {Adamo}, A. and {Oguri}, M. and {Sharon}, K. and {Bradley}, L.~D. and {Christensen}, L. and {Claeyssens}, A. and {Richard}, J. and {Abdurro'uf} and {Bauer}, F.~E. and {Bergamini}, P. and {Bolamperti}, A. and {Brada{\v{c}}}, M. and {Calura}, F. and {Coe}, D. and {Diego}, J.~M. and {Grillo}, C. and {Hsiao}, T.~Y.-Y. and {Inoue}, A.~K. and {Fujimoto}, S. and {Lombardi}, M. and {Meneghetti}, M. and {Resseguier}, T. and {Ricotti}, M. and {Rosati}, P. and {Welch}, B. and {Windhorst}, R.~A. and {Xu}, X. and {Zackrisson}, E. and {Zanella}, A. and {Zitrin}, A.},
        title = "{JWST spectroscopic confirmation of the Cosmic Gems arc at z = 9.625: Insights into the small-scale structure of a post-burst system}",
      journal = {\aap},
         year = 2026,
        month = jan,
       volume = {705},
          eid = {A173},
        pages = {A173},
          doi = {10.1051/0004-6361/202556574},
archivePrefix = {arXiv},
       eprint = {2507.18705},
 primaryClass = {astro-ph.GA},
       adsurl = {https://ui.adsabs.harvard.edu/abs/2026A&A...705A.173M}
}

@ARTICLE{2025A&A...704A..97M,
       author = {{Mini}, I. and {Meneghetti}, M. and {Messa}, M. and {Moscardini}, L. and {Vanzella}, E. and {Bergamini}, P. and {Rosati}, P. and {Zanella}, A.},
        title = "{Image simulations of highly magnified clumpy galaxies}",
      journal = {\aap},
         year = 2025,
        month = dec,
       volume = {704},
          eid = {A97},
        pages = {A97},
          doi = {10.1051/0004-6361/202555682},
archivePrefix = {arXiv},
       eprint = {2512.02114},
 primaryClass = {astro-ph.GA},
       adsurl = {https://ui.adsabs.harvard.edu/abs/2025A&A...704A..97M}
}

@ARTICLE{2025arXiv251107542A,
       author = {{Atek}, Hakim and {Chisholm}, John and {Kokorev}, Vasily and {Endsley}, Ryan and {Pan}, Richard and {Furtak}, Lukas and {Chemerynska}, Iryna and {Richard}, Johan and {Claeyssens}, Ad{\'e}la{\"\i}de and {Oesch}, Pascal and {Fujimoto}, Seiji and {Naidu}, Rohan and {Korber}, Damien and {Schaerer}, Daniel and {Blaizot}, Jeremy and {Rosdahl}, Joki and {Adamo}, Angela and {Asada}, Yoshihisa and {Basu}, Arghyadeep and {Beauchesne}, Benjamin and {Berg}, Danielle and {Bezanson}, Rachel and {Bouwens}, Rychard and {Brammer}, Gabriel and {Dessauges-Zavadsky}, Miroslava and {Ellien}, Ama{\"e}l and {Ezziati}, Meriam and {Fei}, Qinyue and {Goovaerts}, Ilias and {Heurtier}, Sylvain and {Hsiao}, Tiger Yu-Yang and {Jecmen}, Michelle and {Khullar}, Gourav and {Kneib}, Jean-Paul and {Labb{\'e}}, Ivo and {Leclercq}, Floriane and {Marques-Chaves}, Rui and {Mason}, Charlotte and {McQuinn}, Kristen B.~W. and {Mu{\~n}oz}, Julian B. and {Natarajan}, Priyamvada and {Saldana-Lopez}, Alberto and {Stephenson}, Mabel G. and {Trebitsch}, Maxime and {Volonteri}, Marta and {Weibel}, Andrea and {Zitrin}, Adi},
        title = "{JWST's GLIMPSE: an overview of the deepest probe of early galaxy formation and cosmic reionization}",
      journal = {arXiv e-prints},
         year = 2025,
        month = nov,
          eid = {arXiv:2511.07542},
        pages = {arXiv:2511.07542},
          doi = {10.48550/arXiv.2511.07542},
archivePrefix = {arXiv},
       eprint = {2511.07542},
 primaryClass = {astro-ph.GA},
       adsurl = {https://ui.adsabs.harvard.edu/abs/2025arXiv251107542A}
}

@TECHREPORT{2025jwst.rept.9225S,
       author = {{Sunnquist}, Ben and {Boyer}, Martha and {Brooks}, Brian and {Canipe}, Alicia and {Hilbert}, Bryan and {Rest}, Armin},
        title = "{Version 4 of the NIRCam Wisp Templates: Wisp Characterization, Stability, and Validation Testing}",
  institution = {STScI},
         year = 2025,
       number = {Technical Report JWST-STScI-009225},
 howpublished = {Technical Report JWST-STScI-009225, 12 pages},
       adsurl = {https://ui.adsabs.harvard.edu/abs/2025jwst.rept.9225S}
}

@ARTICLE{2024MNRAS.531..997C,
       author = {{Cullen}, F. and {McLeod}, D.~J. and {McLure}, R.~J. and {Dunlop}, J.~S. and {Donnan}, C.~T. and {Carnall}, A.~C. and {Keating}, L.~C. and {Magee}, D. and {Arellano-Cordova}, K.~Z. and {Bowler}, R.~A.~A. and {Begley}, R. and {Flury}, S.~R. and {Hamadouche}, M.~L. and {Stanton}, T.~M.},
        title = "{The ultraviolet continuum slopes of high-redshift galaxies: evidence for the emergence of dust-free stellar populations at z > 10}",
      journal = {\mnras},
         year = 2024,
        month = jun,
       volume = {531},
       number = {1},
        pages = {997-1020},
          doi = {10.1093/mnras/stae1211},
archivePrefix = {arXiv},
       eprint = {2311.06209},
 primaryClass = {astro-ph.GA},
       adsurl = {https://ui.adsabs.harvard.edu/abs/2024MNRAS.531..997C}
}

@ARTICLE{2025ApJ...989...46F,
       author = {{Fujimoto}, Seiji and {Naidu}, Rohan P. and {Chisholm}, John and {Atek}, Hakim and {Endsley}, Ryan and {Kokorev}, Vasily and {Furtak}, Lukas J. and {Pan}, Richard and {Liu}, Boyuan and {Bromm}, Volker and {Venditti}, Alessandra and {Visbal}, Eli and {Sarmento}, Richard and {Weibel}, Andrea and {Oesch}, Pascal A. and {Brammer}, Gabriel and {Schaerer}, Daniel and {Adamo}, Angela and {Berg}, Danielle A. and {Bezanson}, Rachel and {Bouwens}, Rychard and {Chemerynska}, Iryna and {Claeyssens}, Ad{\'e}la{\"\i}de and {Dessauges-Zavadsky}, Miroslava and {Frebel}, Anna and {Korber}, Damien and {Labbe}, Ivo and {Marques-Chaves}, Rui and {Matthee}, Jorryt and {McQuinn}, Kristen B.~W. and {Mu{\~n}oz}, Julian B. and {Natarajan}, Priyamvada and {Saldana-Lopez}, Alberto and {Suess}, Katherine A. and {Volonteri}, Marta and {Zitrin}, Adi},
        title = "{GLIMPSE: An Ultrafaint ≃{}10$^{5}$ M$_{{\ensuremath{\odot}}}$ Pop III Galaxy Candidate and First Constraints on the Pop III UV Luminosity Function at z ≃ 6─7}",
      journal = {\apj},
         year = 2025,
        month = aug,
       volume = {989},
       number = {1},
          eid = {46},
        pages = {46},
          doi = {10.3847/1538-4357/ade9a1},
archivePrefix = {arXiv},
       eprint = {2501.11678},
 primaryClass = {astro-ph.GA},
       adsurl = {https://ui.adsabs.harvard.edu/abs/2025ApJ...989...46F}
}

@ARTICLE{2025ApJ...992...63W,
       author = {{Whitler}, Lily and {Stark}, Daniel P. and {Topping}, Michael W. and {Robertson}, Brant and {Rieke}, Marcia and {Hainline}, Kevin N. and {Endsley}, Ryan and {Chen}, Zuyi and {Baker}, William M. and {Bhatawdekar}, Rachana and {Bunker}, Andrew J. and {Carniani}, Stefano and {Charlot}, St{\'e}phane and {Chevallard}, Jacopo and {Curtis-Lake}, Emma and {Egami}, Eiichi and {Eisenstein}, Daniel J. and {Helton}, Jakob M. and {Ji}, Zhiyuan and {Johnson}, Benjamin D. and {P{\'e}rez-Gonz{\'a}lez}, Pablo G. and {Rinaldi}, Pierluigi and {Tacchella}, Sandro and {Williams}, Christina C. and {Willmer}, Christopher N.~A. and {Willott}, Chris and {Witstok}, Joris},
        title = "{The z {\ensuremath{\gtrsim}} 9 Galaxy UV Luminosity Function from the JWST Advanced Deep Extragalactic Survey: Insights into Early Galaxy Evolution and Reionization}",
      journal = {\apj},
         year = 2025,
        month = oct,
       volume = {992},
       number = {1},
          eid = {63},
        pages = {63},
          doi = {10.3847/1538-4357/adfddc},
archivePrefix = {arXiv},
       eprint = {2501.00984},
 primaryClass = {astro-ph.GA},
       adsurl = {https://ui.adsabs.harvard.edu/abs/2025ApJ...992...63W}
}

@ARTICLE{2008ApJ...686.1503B,
       author = {{Brammer}, Gabriel B. and {van Dokkum}, Pieter G. and {Coppi}, Paolo},
        title = "{EAZY: A Fast, Public Photometric Redshift Code}",
      journal = {\apj},
         year = 2008,
        month = oct,
       volume = {686},
       number = {2},
        pages = {1503-1513},
          doi = {10.1086/591786},
archivePrefix = {arXiv},
       eprint = {0807.1533},
 primaryClass = {astro-ph},
       adsurl = {https://ui.adsabs.harvard.edu/abs/2008ApJ...686.1503B}
}

@ARTICLE{2021ApJS..254...22J,
       author = {{Johnson}, Benjamin D. and {Leja}, Joel and {Conroy}, Charlie and {Speagle}, Joshua S.},
        title = "{Stellar Population Inference with Prospector}",
      journal = {\apjs},
         year = 2021,
        month = jun,
       volume = {254},
       number = {2},
          eid = {22},
        pages = {22},
          doi = {10.3847/1538-4365/abef67},
archivePrefix = {arXiv},
       eprint = {2012.01426},
 primaryClass = {astro-ph.GA},
       adsurl = {https://ui.adsabs.harvard.edu/abs/2021ApJS..254...22J}
}

@ARTICLE{2026arXiv260111704D,
       author = {{Diego}, J.~M. and {Palencia}, J.~M. and {Goolsby}, C. and {Conselice}, C.~J. and {Lagattuta}, D.~J. and {Mahler}, G. and {Richard}, J. and {Sharon}, K. and {Williams}, L.~L.~R.},
        title = "{Hedorah, the first yellow supergiant Kaiju star candidate at $z\approx3$ revealed by behind AS1063}",
      journal = {arXiv e-prints},
         year = 2026,
        month = jan,
          eid = {arXiv:2601.11704},
        pages = {arXiv:2601.11704},
          doi = {10.48550/arXiv.2601.11704},
archivePrefix = {arXiv},
       eprint = {2601.11704},
 primaryClass = {astro-ph.GA},
       adsurl = {https://ui.adsabs.harvard.edu/abs/2026arXiv260111704D}
}

@ARTICLE{2024ApJ...972..143C,
       author = {{Castellano}, Marco and {Napolitano}, Lorenzo and {Fontana}, Adriano and {Roberts-Borsani}, Guido and {Treu}, Tommaso and {Vanzella}, Eros and {Zavala}, Jorge A. and {Arrabal Haro}, Pablo and {Calabr{\`o}}, Antonello and {Llerena}, Mario and {Mascia}, Sara and {Merlin}, Emiliano and {Paris}, Diego and {Pentericci}, Laura and {Santini}, Paola and {Bakx}, Tom J.~L.~C. and {Bergamini}, Pietro and {Cupani}, Guido and {Dickinson}, Mark and {Filippenko}, Alexei V. and {Glazebrook}, Karl and {Grillo}, Claudio and {Kelly}, Patrick L. and {Malkan}, Matthew A. and {Mason}, Charlotte A. and {Morishita}, Takahiro and {Nanayakkara}, Themiya and {Rosati}, Piero and {Sani}, Eleonora and {Wang}, Xin and {Yoon}, Ilsang},
        title = "{JWST NIRSpec Spectroscopy of the Remarkable Bright Galaxy GHZ2/GLASS-z12 at Redshift 12.34}",
      journal = {\apj},
         year = 2024,
        month = sep,
       volume = {972},
       number = {2},
          eid = {143},
        pages = {143},
          doi = {10.3847/1538-4357/ad5f88},
archivePrefix = {arXiv},
       eprint = {2403.10238},
 primaryClass = {astro-ph.GA},
       adsurl = {https://ui.adsabs.harvard.edu/abs/2024ApJ...972..143C}
}

@ARTICLE{2010ApJ...719.1168E,
       author = {{Erb}, Dawn K. and {Pettini}, Max and {Shapley}, Alice E. and {Steidel}, Charles C. and {Law}, David R. and {Reddy}, Naveen A.},
        title = "{Physical Conditions in a Young, Unreddened, Low-metallicity Galaxy at High Redshift}",
      journal = {\apj},
         year = 2010,
        month = aug,
       volume = {719},
       number = {2},
        pages = {1168-1190},
          doi = {10.1088/0004-637X/719/2/1168},
archivePrefix = {arXiv},
       eprint = {1006.5456},
 primaryClass = {astro-ph.CO},
       adsurl = {https://ui.adsabs.harvard.edu/abs/2010ApJ...719.1168E}
}

@ARTICLE{2006AJ....132..926C,
       author = {{Coe}, Dan and {Ben{\'\i}tez}, Narciso and {S{\'a}nchez}, Sebasti{\'a}n F. and {Jee}, Myungkook and {Bouwens}, Rychard and {Ford}, Holland},
        title = "{Galaxies in the Hubble Ultra Deep Field. I. Detection, Multiband Photometry, Photometric Redshifts, and Morphology}",
      journal = {\aj},
         year = 2006,
        month = aug,
       volume = {132},
       number = {2},
        pages = {926-959},
          doi = {10.1086/505530},
archivePrefix = {arXiv},
       eprint = {astro-ph/0605262},
 primaryClass = {astro-ph},
       adsurl = {https://ui.adsabs.harvard.edu/abs/2006AJ....132..926C}
}

@ARTICLE{2010ApJ...712..833C,
       author = {{Conroy}, Charlie and {Gunn}, James E.},
        title = "{The Propagation of Uncertainties in Stellar Population Synthesis Modeling. III. Model Calibration, Comparison, and Evaluation}",
      journal = {\apj},
         year = 2010,
        month = apr,
       volume = {712},
       number = {2},
        pages = {833-857},
          doi = {10.1088/0004-637X/712/2/833},
archivePrefix = {arXiv},
       eprint = {0911.3151},
 primaryClass = {astro-ph.CO},
       adsurl = {https://ui.adsabs.harvard.edu/abs/2010ApJ...712..833C}
}

@software{2016ascl.soft09011B,
       author = {{Bradley}, Larry and {Sipocz}, Brigitta and {Robitaille}, Thomas and {Tollerud}, Erik and {Deil}, Christoph and {Vin{\'\i}cius}, Z{\`e} and {Barbary}, Kyle and {G{\"u}nther}, Hans Moritz and {Bostroem}, Azalee and {Droettboom}, Michael and {Bray}, Erik and {Bratholm}, Lars Andersen and {Pickering}, T.~E. and {Craig}, Matt and {Pascual}, Sergio and {Greco}, Johnny and {Donath}, Axel and {Kerzendorf}, Wolfgang and {Littlefair}, Stuart and {Barentsen}, Geert and {D'Eugenio}, Francesco and {Weaver}, Benjamin Alan},
        title = "{Photutils: Photometry tools}",
 howpublished = {Astrophysics Source Code Library, record ascl:1609.011},
         year = 2016,
        month = sep,
          eid = {ascl:1609.011},
archivePrefix = {ascl},
       eprint = {1609.011},
       adsurl = {https://ui.adsabs.harvard.edu/abs/2016ascl.soft09011B}
}

@ARTICLE{2026arXiv260106015Y,
       author = {{Yanagisawa}, Hiroto and {Ouchi}, Masami and {Golubchik}, Miriam and {Oguri}, Masamune and {Fujimoto}, Seiji and {Kokorev}, Vasily and {Brammer}, Gabriel and {Sun}, Fengwu and {Nakane}, Minami and {Harikane}, Yuichi and {Umeda}, Hiroya and {Akins}, Hollis B. and {Atek}, Hakim and {Bauer}, Franz E. and {Brada{\v{c}}}, Maru{\v{s}}a and {Chisholm}, John and {Coe}, Dan and {Diego}, Jose M. and {Ferguson}, Henry C. and {Finkelstein}, Steven L. and {Furtak}, Lukas J. and {Inayoshi}, Kohei and {Koekemoer}, Anton M. and {Matthee}, Jorryt and {Naidu}, Rohan P. and {Ono}, Yoshiaki and {Pan}, Richard and {Richard}, Johan and {Robbins}, Luke and {Willott}, Chris and {Zitrin}, Adi and {Amor{\'\i}n}, Ricardo O. and {Bradley}, Larry D. and {Bromm}, Volker and {Conselice}, Christopher J. and {Dayal}, Pratika and {Kartaltepe}, Jeyhan S. and {Lopes}, Paulo A.~A. and {Lucas}, Ray A. and {Magdis}, Georgios E. and {Martis}, Nicholas S. and {Papovich}, Casey and {Schaerer}, Daniel and {Valentino}, Francesco and {Vanzella}, Eros and {Allingham}, Joseph F.~V. and {Grogin}, Norman A. and {Gonz{\'a}lez-Otero}, Mauro and {Ricotti}, Massimo and {Windhorst}, Rogier A.},
        title = "{VENUS: Two Faint Little Red Dots Separated by $\sim70\,\mathrm{pc}$ Hidden in a Single Lensed Galaxy at $z\sim7$}",
      journal = {arXiv e-prints},
         year = 2026,
        month = jan,
          eid = {arXiv:2601.06015},
        pages = {arXiv:2601.06015},
          doi = {10.48550/arXiv.2601.06015},
archivePrefix = {arXiv},
       eprint = {2601.06015},
 primaryClass = {astro-ph.GA},
       adsurl = {https://ui.adsabs.harvard.edu/abs/2026arXiv260106015Y}
}

@ARTICLE{2022ApJ...935..167A,
       author = {{Astropy Collaboration} and {Price-Whelan}, Adrian M. and {Lim}, Pey Lian and {Earl}, Nicholas and {Starkman}, Nathaniel and {Bradley}, Larry and {Shupe}, David L. and {Patil}, Aarya A. and {Corrales}, Lia and {Brasseur}, C.~E. and {N{\"o}the}, Maximilian and {Donath}, Axel and {Tollerud}, Erik and {Morris}, Brett M. and {Ginsburg}, Adam and {Vaher}, Eero and {Weaver}, Benjamin A. and {Tocknell}, James and {Jamieson}, William and {van Kerkwijk}, Marten H. and {Robitaille}, Thomas P. and {Merry}, Bruce and {Bachetti}, Matteo and {G{\"u}nther}, H. Moritz and {Aldcroft}, Thomas L. and {Alvarado-Montes}, Jaime A. and {Archibald}, Anne M. and {B{\'o}di}, Attila and {Bapat}, Shreyas and {Barentsen}, Geert and {Baz{\'a}n}, Juanjo and {Biswas}, Manish and {Boquien}, M{\'e}d{\'e}ric and {Burke}, D.~J. and {Cara}, Daria and {Cara}, Mihai and {Conroy}, Kyle E. and {Conseil}, Simon and {Craig}, Matthew W. and {Cross}, Robert M. and {Cruz}, Kelle L. and {D'Eugenio}, Francesco and {Dencheva}, Nadia and {Devillepoix}, Hadrien A.~R. and {Dietrich}, J{\"o}rg P. and {Eigenbrot}, Arthur Davis and {Erben}, Thomas and {Ferreira}, Leonardo and {Foreman-Mackey}, Daniel and {Fox}, Ryan and {Freij}, Nabil and {Garg}, Suyog and {Geda}, Robel and {Glattly}, Lauren and {Gondhalekar}, Yash and {Gordon}, Karl D. and {Grant}, David and {Greenfield}, Perry and {Groener}, Austen M. and {Guest}, Steve and {Gurovich}, Sebastian and {Handberg}, Rasmus and {Hart}, Akeem and {Hatfield-Dodds}, Zac and {Homeier}, Derek and {Hosseinzadeh}, Griffin and {Jenness}, Tim and {Jones}, Craig K. and {Joseph}, Prajwel and {Kalmbach}, J. Bryce and {Karamehmetoglu}, Emir and {Ka{\l}uszy{\'n}ski}, Miko{\l}aj and {Kelley}, Michael S.~P. and {Kern}, Nicholas and {Kerzendorf}, Wolfgang E. and {Koch}, Eric W. and {Kulumani}, Shankar and {Lee}, Antony and {Ly}, Chun and {Ma}, Zhiyuan and {MacBride}, Conor and {Maljaars}, Jakob M. and {Muna}, Demitri and {Murphy}, N.~A. and {Norman}, Henrik and {O'Steen}, Richard and {Oman}, Kyle A. and {Pacifici}, Camilla and {Pascual}, Sergio and {Pascual-Granado}, J. and {Patil}, Rohit R. and {Perren}, Gabriel I. and {Pickering}, Timothy E. and {Rastogi}, Tanuj and {Roulston}, Benjamin R. and {Ryan}, Daniel F. and {Rykoff}, Eli S. and {Sabater}, Jose and {Sakurikar}, Parikshit and {Salgado}, Jes{\'u}s and {Sanghi}, Aniket and {Saunders}, Nicholas and {Savchenko}, Volodymyr and {Schwardt}, Ludwig and {Seifert-Eckert}, Michael and {Shih}, Albert Y. and {Jain}, Anany Shrey and {Shukla}, Gyanendra and {Sick}, Jonathan and {Simpson}, Chris and {Singanamalla}, Sudheesh and {Singer}, Leo P. and {Singhal}, Jaladh and {Sinha}, Manodeep and {Sip{\H{o}}cz}, Brigitta M. and {Spitler}, Lee R. and {Stansby}, David and {Streicher}, Ole and {{\v{S}}umak}, Jani and {Swinbank}, John D. and {Taranu}, Dan S. and {Tewary}, Nikita and {Tremblay}, Grant R. and {de Val-Borro}, Miguel and {Van Kooten}, Samuel J. and {Vasovi{\'c}}, Zlatan and {Verma}, Shresth and {de Miranda Cardoso}, Jos{\'e} Vin{\'\i}cius and {Williams}, Peter K.~G. and {Wilson}, Tom J. and {Winkel}, Benjamin and {Wood-Vasey}, W.~M. and {Xue}, Rui and {Yoachim}, Peter and {Zhang}, Chen and {Zonca}, Andrea and {Astropy Project Contributors}},
        title = "{The Astropy Project: Sustaining and Growing a Community-oriented Open-source Project and the Latest Major Release (v5.0) of the Core Package}",
      journal = {\apj},
         year = 2022,
        month = aug,
       volume = {935},
       number = {2},
          eid = {167},
        pages = {167},
          doi = {10.3847/1538-4357/ac7c74},
archivePrefix = {arXiv},
       eprint = {2206.14220},
 primaryClass = {astro-ph.IM},
       adsurl = {https://ui.adsabs.harvard.edu/abs/2022ApJ...935..167A}
}

@ARTICLE{1983ApJ...266..713O,
       author = {{Oke}, J.~B. and {Gunn}, J.~E.},
        title = "{Secondary standard stars for absolute spectrophotometry.}",
      journal = {\apj},
         year = 1983,
        month = mar,
       volume = {266},
        pages = {713-717},
          doi = {10.1086/160817},
       adsurl = {https://ui.adsabs.harvard.edu/abs/1983ApJ...266..713O}
}

@ARTICLE{2016ARA&A..54..761S,
       author = {{Stark}, Daniel P.},
        title = "{Galaxies in the First Billion Years After the Big Bang}",
      journal = {\araa},
         year = 2016,
        month = sep,
       volume = {54},
        pages = {761-803},
          doi = {10.1146/annurev-astro-081915-023417},
       adsurl = {https://ui.adsabs.harvard.edu/abs/2016ARA&A..54..761S}
}

@ARTICLE{2024ApJ...976..193R,
       author = {{Roberts-Borsani}, Guido and {Treu}, Tommaso and {Shapley}, Alice and {Fontana}, Adriano and {Pentericci}, Laura and {Castellano}, Marco and {Morishita}, Takahiro and {Bergamini}, Pietro and {Rosati}, Piero},
        title = "{Between the Extremes: A JWST Spectroscopic Benchmark for High-redshift Galaxies Using {\ensuremath{\sim}}500 Confirmed Sources at z {\ensuremath{\geq}} 5}",
      journal = {\apj},
         year = 2024,
        month = dec,
       volume = {976},
       number = {2},
          eid = {193},
        pages = {193},
          doi = {10.3847/1538-4357/ad85d3},
archivePrefix = {arXiv},
       eprint = {2403.07103},
 primaryClass = {astro-ph.GA},
       adsurl = {https://ui.adsabs.harvard.edu/abs/2024ApJ...976..193R}
}

@ARTICLE{2023ApJ...943....2W,
       author = {{Welch}, Brian and {Coe}, Dan and {Zitrin}, Adi and {Diego}, Jose M. and {Windhorst}, Rogier and {Mandelker}, Nir and {Vanzella}, Eros and {Ravindranath}, Swara and {Zackrisson}, Erik and {Florian}, Michael and {Bradley}, Larry and {Sharon}, Keren and {Brada{\v{c}}}, Maru{\v{s}}a and {Rigby}, Jane and {Frye}, Brenda and {Fujimoto}, Seiji},
        title = "{RELICS: Small-scale Star Formation in Lensed Galaxies at z = 6-10}",
      journal = {\apj},
         year = 2023,
        month = jan,
       volume = {943},
       number = {1},
          eid = {2},
        pages = {2},
          doi = {10.3847/1538-4357/aca8a8},
archivePrefix = {arXiv},
       eprint = {2207.03532},
 primaryClass = {astro-ph.GA},
       adsurl = {https://ui.adsabs.harvard.edu/abs/2023ApJ...943....2W}
}

@ARTICLE{2024Natur.632..513A,
       author = {{Adamo}, Angela and {Bradley}, Larry D. and {Vanzella}, Eros and {Claeyssens}, Ad{\'e}la{\"\i}de and {Welch}, Brian and {Diego}, Jose M. and {Mahler}, Guillaume and {Oguri}, Masamune and {Sharon}, Keren and {Abdurro'uf} and {Hsiao}, Tiger Yu-Yang and {Xu}, Xinfeng and {Messa}, Matteo and {Lassen}, Augusto E. and {Zackrisson}, Erik and {Brammer}, Gabriel and {Coe}, Dan and {Kokorev}, Vasily and {Ricotti}, Massimo and {Zitrin}, Adi and {Fujimoto}, Seiji and {Inoue}, Akio K. and {Resseguier}, Tom and {Rigby}, Jane R. and {Jim{\'e}nez-Teja}, Yolanda and {Windhorst}, Rogier A. and {Hashimoto}, Takuya and {Tamura}, Yoichi},
        title = "{Bound star clusters observed in a lensed galaxy 460 Myr after the Big Bang}",
      journal = {\nat},
         year = 2024,
        month = aug,
       volume = {632},
       number = {8025},
        pages = {513-516},
          doi = {10.1038/s41586-024-07703-7},
archivePrefix = {arXiv},
       eprint = {2401.03224},
 primaryClass = {astro-ph.GA},
       adsurl = {https://ui.adsabs.harvard.edu/abs/2024Natur.632..513A}
}

@article{
doi:10.1126/science.ady9404,
author = {Yuduo Guo  and Hao Zhang  and Mingyu Li  and Fujiang Yu  and Yunjing Wu  and Yuhan Hao  and Song Huang  and Yongming Liang  and Xiaojing Lin  and Xinyang Li  and Jiamin Wu  and Zheng Cai  and Qionghai Dai },
title = {Deeper detection limits in astronomical imaging using self-supervised spatiotemporal denoising},
journal = {Science},
volume = {392},
number = {6797},
pages = {eady9404},
year = {2026},
doi = {10.1126/science.ady9404},
URL = {https://www.science.org/doi/abs/10.1126/science.ady9404},
eprint = {https://www.science.org/doi/pdf/10.1126/science.ady9404}}

@article{2003A&A...397..527S,
  author  = {{Schaerer}, D.},
  title   = "{The transition from Population III to normal galaxies: Lyalpha and He II emission and the ionising properties of high redshift starburst galaxies}",
  journal = {Astronomy and Astrophysics},
  year    = {2003},
  volume  = {397},
  pages   = {527--538},
  doi     = {10.1051/0004-6361:20021525},
  eprint  = {astro-ph/0210462}
}

@article{2019MNRAS.488.3492S,
  author  = {{Senchyna}, P. and {Stark}, D.~P. and {Chevallard}, J. and {Charlot}, S. and {Jones}, T. and {Vidal-Garc{\'i}a}, A.},
  title   = "{Extremely metal-poor galaxies with HST/COS: laboratories for models of low-metallicity massive stars and high-redshift galaxies}",
  journal = {Monthly Notices of the Royal Astronomical Society},
  year    = {2019},
  volume  = {488},
  pages   = {3492--3506},
  doi     = {10.1093/mnras/stz1907},
  eprint  = {1904.01615}
}

@article{2022ApJ...941..153T,
  author  = {{Topping}, M.~W. and {Stark}, D.~P. and {Endsley}, R. and {Plat}, A. and {Whitler}, L. and {Chen}, Z. and {Charlot}, S.},
  title   = "{Searching for Extremely Blue UV Continuum Slopes at z = 7--11 in JWST/NIRCam Imaging: Implications for Stellar Metallicity and Ionizing Photon Escape in Early Galaxies}",
  journal = {The Astrophysical Journal},
  year    = {2022},
  volume  = {941},
  number  = {2},
  eid     = {153},
  pages   = {153},
  doi     = {10.3847/1538-4357/aca522},
  eprint  = {2208.01610}
}

@article{2023A&A...678A.173V,
  author  = {{Vanzella}, E. and {Claeyssens}, A. and {Welch}, B. and {Adamo}, A. and {Zackrisson}, E. and {Bergamini}, P. and {Diego}, J.~M. and {Messa}, M. and {Meneghetti}, M. and {Zitrin}, A. and {Coe}, D. and {Bradley}, L.~D. and others},
  title   = "{JWST/NIRSpec spectroscopy of a strongly lensed stellar complex at z = 6.629: a possible Population III host}",
  journal = {Astronomy and Astrophysics},
  year    = {2023},
  volume  = {678},
  eid     = {A173},
  pages   = {A173},
  doi     = {10.1051/0004-6361/202346981},
  eprint  = {2305.14413}
}

@article{2023MNRAS.525.5328T,
  author  = {{Trussler}, J.~A.~A. and {Conselice}, C.~J. and {Adams}, N.~J. and {Maiolino}, R. and {Nakajima}, K. and {Zackrisson}, E. and {Austin}, D. and {Ferreira}, L. and {Harvey}, T.},
  title   = "{On the observability and identification of Population III galaxies with JWST}",
  journal = {Monthly Notices of the Royal Astronomical Society},
  year    = {2023},
  volume  = {525},
  pages   = {5328--5352},
  doi     = {10.1093/mnras/stad2553},
  eprint  = {2211.02038}
}

@article{2024A&A...687A..67M,
  author  = {{Maiolino}, R. and {Scholtz}, J. and {Curtis-Lake}, E. and {Carniani}, S. and {Bunker}, A.~J. and {Baker}, W.~M. and {Curti}, M. and {Witstok}, J. and {D'Eugenio}, F. and {Eisenstein}, D.~J. and others},
  title   = "{JADES. Possible Population III signatures at z = 10.6 in the halo of GN-z11}",
  journal = {Astronomy and Astrophysics},
  year    = {2024},
  volume  = {687},
  eid     = {A67},
  pages   = {A67},
  doi     = {10.1051/0004-6361/202347087},
  eprint  = {2306.00953}
}

@article{2024MNRAS.529.3301T,
  author  = {{Topping}, M.~W. and {Stark}, D.~P. and {Senchyna}, P. and {Plat}, A. and {Zitrin}, A. and {Endsley}, R. and {Charlot}, S. and {Furtak}, L.~J. and {Maseda}, M.~V. and {Smit}, R. and {Mainali}, R. and {Chevallard}, J. and {Molyneux}, S. and {Rigby}, J.~R.},
  title   = "{Metal-poor star formation at z > 6 with JWST: new insight into hard radiation fields and nitrogen enrichment on 20 pc scales}",
  journal = {Monthly Notices of the Royal Astronomical Society},
  year    = {2024},
  volume  = {529},
  pages   = {3301--3319},
  doi     = {10.1093/mnras/stae807},
  eprint  = {2401.08764}
}
\bibliographystyle{aasjournalv7}

\end{document}